\documentclass[aps,prd,twocolumn,superscriptaddress]{revtex4-1}

\newcommand{\udt}[3]{#1^{#2}_{\phantom{#2}#3}}

\newcommand{\dut}[3]{#1_{#2}^{\phantom{#2}#3}}
\newcommand{\dudt}[4]{#1_{#2\phantom{#3}#4}^{\phantom{#2}#3}}
\newcommand{\udut}[4]{#1^{#2\phantom{#3}#4}_{\phantom{#2}#3}}

\usepackage{amsmath,amssymb,amsfonts,mathtools}
\usepackage{graphicx,epsfig}
\usepackage{hyperref}
\usepackage{bm,color}
\usepackage{tabularx,siunitx}
\usepackage{wasysym}

\begin{document}

\title{Gravitoelectromagnetism, Solar System Test and Weak-Field Solutions in \texorpdfstring{$f(T,B)$}{fTB} Gravity with Observational Constraints}

\author{Gabriel Farrugia}
\email[]{gabriel.farrugia.11@um.edu.mt}

\author{Jackson Levi Said}
\email[]{jackson.said@um.edu.mt}

\author{Andrew Finch}
\email[]{andrew.finch.12@um.edu.mt}
\affiliation{Institute of Space Sciences and Astronomy, University of Malta, Malta}
\affiliation{Department of Physics, University of Malta, Malta}

\date{\today}

\begin{abstract}
Gravitomagnetism characterize phenomena in the weak field limit within the context of rotating systems. These are mainly manifested in the geodetic and Lense-Thirring effects. The geodetic effect describes the precession of the spin of a gyroscope in orbit about a massive static central object, while the Lense-Thirring effect expresses the analogous effect for the precession of the orbit about a rotating source. In this work, we explore these effects in the framework of Teleparallel Gravity and investigate how these effects may impact recent and future missions. We find that teleparallel theories of gravity may have an important impact on these effects which may constrain potential models within these theories.
\end{abstract}

\maketitle

\section{Introduction}
General relativity (GR) has passed numerous observational tests since its inception just over a century ago, confirming its predictive power. The detection of gravitational waves in 2015 \cite{Abbott:2016blz} agreed with the strong field predictions of GR, as does its solar system behaviour \cite{Will:2014kxa}. However, GR requires a large portion of dark matter to explain the dynamics of galaxies \cite{Bertone:2004pz,Navarro:1995iw} and even greater contributions from dark energy to produce current observations of cosmology \cite{peebles:1993}. Given the lack of a concrete theoretical explanation of these phenomena we are motivated to explore the possibility of modifying gravity within the observational context.

There are a myriad of ways in which to consider modified theories of gravity \cite{Sotiriou:2008rp,Capozziello:2011et,Clifton:2011jh}, and to constrain them \cite{Iorio:2018fam,Iorio:2016sqy,Deng:2015sua}. These range from extensions to the standard gravity of GR to more exotic directions. One interesting framework that has gained attention in recent years is that of Teleparallel Gravity (TG). TG is formed by first considering a connection that is not curvature-full, i.e. we consider a connection that is distinct from the regular Levi-Civita connection (which forms the Christoffel symbols). In this way, the gravitational contributions to the metric tensor become a source of torsion rather than curvature. This is achieved by replacing the Levi-Civita connection with its Weitzenb\"{o}ck analog. The Weitzenb\"{o}ck connection is torsion-full while being curvature-less and satisfying the metricity condition \cite{Weitzenbock1923}. Thus, we can construct theories of gravity which express gravitation through torsion rather than curvature. One such theory is the teleparallel equivalent of general relativity (TEGR) which produces the same dynamical equations as GR while being sourced by a different gravitational action, i.e. one that is based on torsion rather than curvature.

TEGR and GR differ in their Lagrangians by a boundary term that plays an important role in the extensions of these theories \cite{Cai:2015emx,Krssak:2018ywd,Maluf:2013gaa}. The boundary term naturally appears in GR due to the appearance of second-order derivatives in the Lagrangian \cite{Nakahara:2003nw,ortin2004gravity}, which is the core difference between GR and TEGR at the level of the Action.

In fact, this boundary term is the source of the fourth-order contributions to $f(R)$ theories of gravity. For this reason, TG features a weakened Lovelock theorem \cite{Lovelock:1971yv,Gonzalez:2015sha,Bahamonde:2019shr}, which as a direct result means that many more theories of gravity can be constructed that are generally second-order in their field equation derivatives. It is for this reason that TG is very interesting because it organically avoids Gauss-Ostrogradski ghosts in so many contexts. The TEGR Lagrangian can be immediately generalize to produce $f(T)$ gravity \cite{Ferraro:2006jd,Ferraro:2008ey,Bengochea:2008gz,Linder:2010py,Chen:2010va} in the same way that the Einstein-Hilbert action leads to $f(R)$ gravity. A number of $f(T)$ gravity models have shown promising results in the solar system regime \cite{Farrugia:2016xcw,Iorio:2012cm,Ruggiero:2015oka,Deng:2018ncg}, as well as in the galactic \cite{Finch:2018gkh} and cosmological regimes \cite{Cai:2015emx,Nesseris:2013jea,Farrugia:2016qqe}. Of particular interest is its effect on weak lensing in galaxy-galaxy surveys \cite{Chen:2019ftv}. However, to fully incorporate $f(R)$ gravity, we must consider $f(T,B)$ gravity where $B$ represents a boundary term that appears as the difference between the Ricci scalar and the torsion scalar (and will be discussed in more detail in \S.\ref{TG_rev}).

Gravitomagnetic tests offer offer an ideal vehicle to probe the rotational behaviour of theories of gravity in their weak field limits \cite{Cohen:1993jm,hartle:2003,Mashhoon:2019jkq,Lichtenegger:2002af,Mashhoon:2003ax,Iorio:2017puo}. In fact, gravitomagnetic effects are the result of mass currents appearing in the weak field limit of GR where the Einstein field equations take on a form reminiscent of Maxwell's equations \cite{misner1973gravitation} (and do not involve actual electromagnetic effects). These effects emerge as a result of a rotating source or observer in a system, which both give independent contributions to the overall observational effect. For the case where an orbiting observer is moving about a stationary source, Geodetic effects emerge \cite{misner1973gravitation} where a vector will exhibit precession due to the background spacetime being curved. This is the general relativistic analog of the well-known Thomas precession exhibited in special relativity \cite{Thomas:1926dy}. Another closely related relativistic precession phenomenon is that of the Lense-Thirring effect (or frame-dragging effect) \cite{1918PhyZ...19..156L} where the neighbourhood of a large rotating source causes precession in nearby gyroscopes. While independent, these effects are often observed as a combined observable phenomenon such as in the Earth-Moon system about the Sun \cite{PhysRevLett.4.215}, where the precession of the Moon's perigee is caused by this phenomenon \cite{PhysRevLett.58.1062,PhysRevLett.61.2643,Dickey482}.

Motivated by the Gravity Probe B experiment \cite{Gravity_Probe_B_1}, there have been a number of investigations into the behaviour and predictions of modified theories of gravity \cite{Tartaglia:1998rh,Said:2014lua,Finch:2016bum,Matsuno:2009nz}. However, the accuracy of this experiment is not enough to adequately differentiate between competing models of gravity. There have also been other experimental efforts such as LAGEOS \cite{Ciufolini:2004rq,Ciufolini:2010zz} which aimed to perform lasers test while in orbit about the Earth. The MGS spacecraft \cite{Iorio_2006,Iorio2010} tested gravitomagnetism effects about Mars, while there have also been tests about the Sun \cite{Iorio2012}. For this reason, there have been a number of ambitious proposals put forward in recent years to further test this relativistic effect and to increase the experimental precision of the observations \cite{Ruggiero:2018jab,Tartaglia:2016jfo,Ruggiero:2002hz,DiVirgilio:2017aev}.

In this work, we explore the gravitomagentic effects of TG through $f(T,B)$ gravity, as well as the classical solar system tests within this context. We do this by first expanding into the weak field limit of the theory and explore both the Geodetic and Lense-Thirring effects separately. We then compare their combined results against the recent observations. The manuscript is divided as follows, in \S.\ref{TG_rev} we briefly review and introduce TG and its $f(T,B)$ gravity extension. In \S.\ref{weak_field_approx}, we explore the weak field regime of $f(T,B)$ gravity and discuss some important properties of the theory in this limit. Perturbations about a statis spherically symmetruc metric are considered in \S.\ref{metric_per_sec}. The core results associated with gravitomagentism and the classical solar system tests are then determined in \S.\ref{ob_const_sec}, while a comparison with observational values is presented in \S.\ref{obs_fitting}. Finally we conclude in \S.\ref{sec:conclusion} with some remarks and a discussion. Throughout the manuscript, the speed of light is not set to unity for comparison purposes in the electrodynamics analysis in \S.\ref{weak_field_approx}.

\section{Teleparallel Gravity and its Extension to \texorpdfstring{$f(T,B)$}{ft} gravity \label{TG_rev}}

Teleparallel Gravity represents a paradigm shift in the way that gravity is expressed where curvature is replaced by torsion through an exchange of the Levi-Civita connection, $\mathring{\Gamma}^{\sigma}_{\mu\nu}$, with its Weitzenb\"{o}ck analog, $\Gamma^{\sigma}_{\mu\nu}$, (we use over-dots to represent quantities determined using the Levi-Civita connection) \cite{ortin2004gravity}. GR expresses curvature through the Levi-Civita which is torsion-less, while the Weitzenb\"{o}ck connection is curvature-less and also satisfies the metricity condition \cite{PhysRevD.19.3524}. In theories based on the Levi-Civita connection, curvature is given a meaningful measure by means of the Riemann tensor on Riemannian manifolds \cite{misner1973gravitation}. This formulation of gravity is retained in most popular modified theories of gravity where gravitation continues to be expressed in terms of curvature of a background geometry. However in TG, irrespective of the form of the metric tensor, the Riemann tensor must vanish since the Weitzenb\"{o}ck connection is curvature-less \cite{Hehl:1994ue}. It is for this reason that TG necessitates a fundamental reformulation of gravitation in order to construct realistic models of gravity.

GR and its variants utilize the metric, $g_{\mu\nu}$, as their fundamental dynamical object, but TG treats this as a derived quantity which emerges from the tetrad, $\udt{e}{a}{\mu}$. The tetrad acts as a soldering agent between the general manifold (Greek indices) and its tangent space (Latin indices) \cite{Aldrovandi:2013wha}. Through this action, the tetrad (and its inverses $\dut{e}{a}{\mu}$) can be used to transform indices between these manifolds
\begin{align}
    g_{\mu\nu}=\udt{e}{a}{\mu}\udt{e}{b}{\nu}\eta_{ab}\,,& &\eta_{ab}=\dut{e}{a}{\mu}\dut{e}{b}{\nu}g_{\mu\nu}\,. \label{eq:metric-tetrad-rel}
\end{align}
Moreover, these tetrads observe orthogonality conditions
\begin{align}\label{tetrad_def}
    \udt{e}{a}{\mu}\dut{e}{b}{\mu}=\delta_b^a\,,& &\udt{e}{a}{\mu}\dut{e}{a}{\nu}=\delta_{\mu}^{\nu}\,,
\end{align}
for internal consistency. The Weitzenb\"{o}ck connection is then defined using the tetrad as \cite{Cai:2015emx,Krssak:2018ywd,Weitzenbock1923,Hehl:1994ue}
\begin{equation}
    \Gamma^{\sigma}_{\mu\nu} := \dut{e}{a}{\sigma}\partial_{\mu}\udt{e}{a}{\nu} + \dut{e}{a}{\sigma}\udt{\omega}{a}{b\mu}\udt{e}{b}{\nu}\,,
\end{equation}
where $\udt{\omega}{a}{b\mu}$ is the inertial spin connection. The Weitzenb\"{o}ck connection is the most general linear affine connection that is both curvature-less and satisfies the metricity condition \cite{Aldrovandi:2013wha}. The appearance of the spin connection is there to retain the covariance of the resulting field equations \cite{Krssak:2015rqa}. This is an issue in TG due to the freedom in the choice of the components of the tetrads, that is, there is an infinite number of tetrads that produce the same metric tensor in Eq.(\ref{tetrad_def}). These different tetrads are related by local Lorentz transformations (LLTs). As a result, the spin connection components take on values to account for the LLT invariance of the underlying theory. Thus, there is a particular choice of frames in which the spin connection components are allowed to be zero \cite{Cai:2015emx}. 

In GR, this issue is hidden in the internal structure of the theory \cite{misner1973gravitation}. Considering the full breadth of LLTs (boosts and rotations), $\udt{\Lambda}{a}{b}$, the spin connection can be represented as $\udt{\omega}{a}{b\mu}=\udt{\Lambda}{a}{c}\partial_{\mu}\dut{\Lambda}{b}{c}$ \cite{Krssak:2018ywd}. Thus, it is the combination of tetrad and an associated spin connection that forms the covariance of TG.

Given a Riemann tensor that measures curvature, we must define a so-called torsion tensor that gives a meaningful measure of torsion, defined as \cite{Cai:2015emx}
\begin{equation}\label{eq:tortens-def}
    \udt{T}{\sigma}{\mu\nu} := 2\Gamma^{\sigma}_{[\mu\nu]}\,,
\end{equation}
where the square brackets represent the anti-symmetry operator. The torsion tensor represents the field strength of TG, and transforms covariantly under both diffeomorphisms and LLTs \cite{Aldrovandi:2013wha}. To formulate a gravitational action, we must define two other quantities. Firstly, consider the contorsion tensor which effectively is the difference between the Levi-Civita and Weitzenb\"{o}ck connections, defined as \cite{Maluf:1995re,ortin2004gravity}
\begin{equation}\label{eq:contorsiondef}
    \udt{K}{\sigma}{\mu\nu} := \mathring{\Gamma}^{\sigma}_{\mu\nu} - \Gamma^{\sigma}_{\mu\nu} = \frac{1}{2}\left(\dudt{T}{\mu}{\sigma}{\nu} + \dudt{T}{\nu}{\sigma}{\mu} - \udt{T}{\sigma}{\mu\nu}\right)\,,
\end{equation}
which plays a crucial role in relating TG results with Levi-Civita connection based theories. Secondly, we also need the superpotential which is defined as \cite{Aldrovandi:2013wha,Maluf:1994ji}
\begin{equation}\label{eq:superpotential_def}
    \dut{S}{a}{\mu\nu} := \udt{K}{\mu\nu}{a} - \dut{e}{a}{\nu}\udt{T}{\alpha\mu}{\alpha} + \dut{e}{a}{\mu} \udt{T}{\alpha\nu}{\alpha}\,.
\end{equation}
This has been shown to potentially relate TG to a gauge current representation of the energy-momentum tensor for gravitation \cite{Aldrovandi:2003pa,Koivisto:2019jra}. Then, by contracting the torsion and superpotential tensors, the torsion scalar can be defined as
\begin{equation}\label{torsion_scalar_def}
    T := \dut{S}{a}{\mu\nu}\udt{T}{a}{\mu\nu}\,,
\end{equation}
which is entirely determine by the Weitzenb\"{o}ck connection along the same vain as the Ricci scalar being determined completely by the Levi-Civita connection. Naturally, the Ricci scalar calculated with the Weitzenb\"{o}ck connection will vanish since it is a measure of curvature. This property in conjunction with the use of the contorsion tensors allows for a relation between the regular Ricci scalar and the torsion scalar defined in Eq.(\ref{torsion_scalar_def}) through \cite{Aldrovandi:2013wha,Cai:2015emx,Krssak:2018ywd}
\begin{equation}
    R = \mathring{R} + T - \frac{2}{e}\partial_{\mu}\left(e\udut{T}{\sigma}{\sigma}{\mu}\right) = 0\,,
\end{equation}
where $R$ is the Ricci scalar calculated using the Weitzenb\"{o}ck connection, $\mathring{R}$ is the standard gravity Ricci scalar determined using the regular Levi-Civita connection, and $e=\text{det}\left(\udt{e}{a}{\mu}\right)=\sqrt{-g}$ is the determinant of the tetrad. Thus, the standard Ricci and torsion scalars turn out to be equivalent up to a total divergence term
\begin{equation}
    \mathring{R} = -T + \frac{2}{e}\partial_{\mu}\left(e\udut{T}{\sigma}{\sigma}{\mu}\right) := -T+B\,,
\end{equation}
where $B=2\mathring{\nabla}_{\mu}\left(\udut{T}{\sigma}{\sigma}{\mu}\right)$ is a total divergence term. This relation guarantees that the ensuing equations of motion will be equivalent. Thus, the TEGR action can be written as \cite{Aldrovandi:2013wha,Cai:2015emx}
\begin{equation}
    S_{\text{TEGR}} = -\frac{1}{2\kappa^2}\int d^4 x\, eT + \int d^4 x\, e\mathcal{L}_m\,,
\end{equation}
where $\kappa^2=8\pi G/c^4$, and $\mathcal{L}_m$ is the matter Lagrangian. This action leads to the equivalent dynamical equations as the Einstein-Hilbert action, but the difference in their Lagrangians means that the fourth-order boundary terms are not necessary to form a covariant theory within the TG context. While this does not effect the TEGR limit, it will influence the possible theories that can be formed in the modified gravity scenario.

Considering the same reasoning that led to $f(\mathring{R})$ gravity \cite{Capozziello:2011et,Sotiriou:2008rp}, the Lagrangian of TEGR can be immediately generalized to $f(T)$ gravity \cite{Ferraro:2006jd,Ferraro:2008ey,Bengochea:2008gz,Linder:2010py,Chen:2010va}. The $f(T)$ gravity setting produces generally second-order field equations in terms of derivatives of the tetrads \cite{Cai:2015emx}. This feature is only possible due to a weakening of Lovelock's theorem in the TG setting \cite{Lovelock:1971yv,Gonzalez:2015sha,Bahamonde:2019shr}. This fact alone guarantees that $f(T)$ gravity will not exhibit Gauss-Ostrogradsky ghosts since it remains second-order. $f(T)$ gravity also shares other properties with TEGR such as its GW polarization signature \cite{Farrugia:2018gyz,Abedi:2017jqx}.

However, to fully encompass the breadth of $f(\mathring{R})$ gravity, we must consider the generalization to $f(T,B)$ gravity which contains as a subset the limit $f(\mathring{R})=f(-T+B)$. Thus, $f(T,B)$ gravity is further generalization of $f(\mathring{R})$ in which the second- and fourth-order contributions to the theory are decoupled \cite{Bahamonde:2015zma}.

In this work, we investigate the gravitomagnetic effects of $f(T,B)$ gravity and its effect on observational constraints of the theory for particular models of this setting \cite{Bahamonde:2015zma,Capozziello:2018qcp,Bahamonde:2016grb,Paliathanasis:2017flf,Farrugia:2018gyz,Bahamonde:2016cul,Wright:2016ayu}. To do this, we need the field equations of the theory, which are determined by a variation of the $f(T,B)$ gravitational Lagrangian density, $ef(T,B)$ to give \cite{Krssak:2015oua,Cai:2015emx,Krssak:2018ywd}
\begin{align}
    \dut{e}{a}{\lambda}&\Box f_B - \dut{e}{a}{\sigma}\nabla^{\lambda}\nabla_{\sigma} f_B + \frac{1}{2} B f_B \dut{e}{a}{\lambda}\nonumber\\
    &+2\dut{S}{a}{\mu\lambda}\left[\partial_{\mu}f_T+\partial_{\mu} f_B\right] + \frac{2}{e} f_T \partial_{\mu}\left(e\dut{S}{a}{\mu\lambda}\right) \nonumber\\
    &- 2 f_T \udt{T}{\sigma}{\mu a}\dut{S}{\sigma}{\lambda\mu} - \frac{1}{2}f\dut{e}{a}{\lambda} = \kappa^2 \dut{\Theta}{a}{\lambda}\,, \label{eq:fTB-fieldeq-tetradform}
\end{align}
where subscripts denote derivatives, and $\dut{\Theta}{\rho}{\nu}$ is the regular energy-momentum tensor. The spin connection is taken to be zero \cite{Bahamonde:2015zma,Capozziello:2018qcp,Bahamonde:2016grb,Farrugia:2018gyz} since this will be a demand in the work that follows. We will revisit this statement at various stages of the analysis to confirm the consistency of the work. Using the contorsion tensor relations, the $f(T,B)$ gravity field can also be represented as 
\begin{align}
&-f_T \mathring{G}_{\mu\nu} + \left(g_{\mu\nu} \Box - \nabla_\mu \nabla_\nu\right)f_B + \frac{1}{2}g_{\mu\nu} \left(Bf_B + Tf_T - f\right) \nonumber \\
&+ 2\dudt{S}{\nu}{\alpha}{\mu} \partial_\alpha\left(f_T + f_B\right) = -\kappa^2 \Theta_{\mu\nu}\,, \label{eq:fTB-fieldeq-spaceindex}
\end{align}
where $\mathring{G}_{\mu\nu}$ is the regular Einstein tensor calculated with the Levi-Civita connection. In this setting, the spin connection depends on the choice of tetrad components and so does not produce independent field equations. However, works exist in the literature that consider this scenario such as Refs.\cite{Krssak:2018ywd,Golovnev:2017dox} where a Palatini approach is considered so that a second set of field equations are produced for the spin connection.

\section{The Weak-Field Approximation \label{weak_field_approx}}

\subsection{The Field Equations}

Linearised gravity offers a relatively simple procedure to examine the weak-field metric for a given source. As the gravitational field is assumed to be weak, the metric can be expressed as a Minkowski background plus a small (first order) correction, $h_{\mu\nu}$. In other words, the metric tensor can be expanded as  
\begin{equation}
g_{\mu\nu} = \eta_{\mu\nu} + h_{\mu\nu},
\end{equation}
with $\left|h_{\mu\nu}\right| \ll 1$. By extension, a similar consideration can be applied for the linearised expansion for the tetrad: a background value $\gamma^{(0)a}_\mu$ which yields the Minkowski metric plus some small correction $\gamma^{(1)a}_\mu$, namely 
\begin{equation}
\udt{e}{a}{\mu} = \gamma^{(0)a}_\mu + \gamma^{(1)a}_\mu,
\end{equation}
with $\left|\gamma^{(1)a}_\mu\right| \ll \left|\gamma^{(0)a}_\mu\right| \sim 1$. Following the methodology considered in Ref.\cite{Farrugia:2018gyz}, the resulting perturbed torsional quantities and field equations can be derived. Through the relation between the metric and the tetrad given in Eq.\eqref{eq:metric-tetrad-rel}, the perturbed quantities are interlinked as
\begin{align}
\eta_{\mu\nu} &= \eta_{ab} \gamma^{(0)a}_\mu \gamma^{(0)b}_\nu, \label{eq:MinkMet-linearisation}\\
h_{\mu\nu} &= \eta_{ab} \left(\gamma^{(0)a}_\mu\gamma^{(1)b}_\nu + \gamma^{(1)a}_\mu \gamma^{(0)b}_\nu\right). \label{eq:GWs-perturbed-metric-1st-order-relation}
\end{align}

Given the equations are constructed in the Weitzenb\"{o}ck gauge $\left(\omega_{ab\mu} = 0\right)$, this imposes a constraint on the behaviour of $\gamma^{(0)a}_\mu$. The spin connection takes the form \cite{Farrugia:2018gyz}
\begin{equation}\label{eq:spin-connection-GW}
\udt{\omega}{a}{b\mu} = -\gamma^{(0)\nu}_b \partial_\mu \gamma^{(0)a}_\nu,
\end{equation}
which when compared to its LLT form reveals that the background tetrad corresponds to the Lorentz matrices. This is expected as this background tetrad represent a trivial frame, one which constructs the Minkowski metric \cite{Aldrovandi:2013wha}. As the spin connection is zero here, the background tetrad reduces to a constant, i.e. to the class of constant Lorentz matrices. For simplicity, the background tetrad can be chosen to be $\gamma^{(0)a}_\mu = \delta^a_\mu$ \footnote{Other works which also make this choice within the Weitzenb{\"{o}}ck gauge appear, for instance, in Refs.\cite{Abedi:2017jqx,Bamba:2013ooa}.}

Under these considerations, the torsion tensor Eq.\eqref{eq:tortens-def} turns out to be a first order quantity in the perturbations 
\begin{equation}\label{eq:TorTens-perturbed}
\udt{T}{a}{\mu\nu} = \partial_\mu \gamma^{(1)a}_\nu - \partial_\nu \gamma^{(1)a}_\mu. \end{equation} 
Consequently, as both the contorsion Eq.\eqref{eq:contorsiondef} and superpotential Eq.\eqref{eq:superpotential_def} tensors are linearly dependent on the torsion tensor, then these are also of at least first order. Ultimately, this implies that the torsion scalar is of at least second order. Observe that this result holds true even if the Weitzenb\"{o}ck gauge is not imposed \cite{Farrugia:2018gyz}.

On the other hand, the boundary term is first order. This is also consistent with the relation $\mathring{R} = -T + B$ as the Ricci scalar is of at least first order. Indeed, the Ricci tensor and Ricci scalar are given to be
\begin{align}
\mathring{R}_{\mu\nu} &= \frac{1}{2}\left(\partial_\rho\partial_\mu \udt{h}{\rho}{\nu} + \partial_\rho\partial_\nu \udt{h}{\rho}{\mu} - \partial_\mu \partial_\nu h - \Box h_{\mu\nu}\right), \\
\mathring{R} &= \partial_\rho\partial_\nu h^{\rho\nu} - \Box h,
\end{align}
where $h \coloneqq \udt{h}{\mu}{\mu}$ represents the trace. It is remarked that from here onwards, indices are raised and lowered with respect to the Minkowski (background) metric. Moreover, the d'Alembert operator reduces to $\Box = \partial_\mu \partial^\mu$. 

The next step would be to extract the perturbed field equations. For simplicity, as both $T$ and $B$ are null at a background level, the gravitational Lagrangian $f(T,B)$ is assumed to be Taylor expandable about these latter values, namely
\begin{align}
f(T,B) &= f(0,0) + f_T(0,0) T + f_B(0,0) B \nonumber \\
&+ \frac{1}{2}f_{TT}(0,0) T^2 + \frac{1}{2}f_{BB}(0,0) B^2 \nonumber \\
&+ f_{TB}(0,0) TB + \text{\dots{}}\, .
\end{align}
Observe that the coefficient $f_T(0,0) \neq 0$ as this corresponds to the effective Newtonian gravitational constant as evident from the field equations Eq.\eqref{eq:fTB-fieldeq-spaceindex} (see for instance Refs.\cite{Zheng:2010am,Capozziello:2011hj} for detailed discussions in the case of $f(T)$ gravity). Under this assumption, the zeroth and first order field equations of $f(T,B)$ gravity Eq.\eqref{eq:fTB-fieldeq-spaceindex} are
\begin{align}
\eta_{\mu\nu}f(0,0) &= 0, \\
f_T(0,0) \mathring{G}_{\mu\nu} - f_{BB}(0,0) \left(\eta_{\mu\nu} \Box - 
\partial_\mu \partial_\nu\right) \mathring{R} &= \kappa^2 \Theta_{\mu\nu}, \label{eq:fTB-first-order}
\end{align}
where the result $\mathring{R} = B$ (which is valid up to this order) has been used, a property which shall be useful in order to simplify the forthcoming equations. The zeroth order equation confirms the absence of a cosmological constant $2\Lambda \equiv f(0,0)$, maintaining consistency with the linearisation regime as the background geometry is Minkowski spacetime. 

As mentioned previously, $f(\mathring{R})$ gravity is a sub-case of $f(T,B)$ gravity. In fact, the resulting perturbed equations Eq.\eqref{eq:fTB-first-order} are practically identical in form to those found in $f(\mathring{R})$ gravity with the only difference being in the form of the coefficients \cite{Capozziello:2007ms,Capozziello:2008rq,Corda:2010zza,Berry:2011pb,Yang:2011cp,Capozziello:2011et,Rizwana:2016qdq,Liang:2017ahj,Gong:2017bru,Gong:2018ybk}. Motivated by this, the same procedure as presented in Refs.\cite{Abedi:2017jqx} shall be followed. 

First, the quantity $\bar{h}_{\mu\nu}$ defined as
\begin{equation}\label{eq:fTB-trace-reversed-def}
h_{\mu\nu} = \bar{h}_{\mu\nu} - \frac{1}{2}\bar{h}\eta_{\mu\nu} + \frac{f_{BB}(0,0)}{f_T(0,0)}\eta_{\mu\nu}R,
\end{equation}
is introduced, with $\bar{h} \coloneqq \udt{\bar{h}}{\mu}{\mu}$. As shown in Refs.\cite{Myung:2016zdl,Liang:2017ahj}, the Lorenz gauge $\partial^\mu \bar{h}_{\mu\nu} = 0$ can be imposed. In this way, the field equations Eq.\eqref{eq:fTB-first-order} take a relatively simple form
\begin{equation}\label{eq:fTB-perturbed-h-fieldeq}
\Box \bar{h}_{\mu\nu} = -\frac{2\kappa^2}{f_T(0,0)}\Theta_{\mu\nu}.
\end{equation} 
The next step is to obtain the form of the perturbed Ricci scalar. Taking the trace of Eq.\eqref{eq:fTB-first-order} yields the relation
\begin{equation}\label{eq:fTB-trace-1st}
f_T(0,0) R + 3 f_{BB}(0,0) \Box \mathring{R} = -\kappa^2\Theta,
\end{equation}
which is of the same form as the Klein-Gordon equation having an effective mass
\begin{equation}\label{eq:fTB-eff-mass}
\mu^2 \equiv -\frac{f_T(0,0)}{3 f_{BB}(0,0)}.
\end{equation}

Depending on the form of the source (and hence of the stres-energy tensor), Eqs.\eqref{eq:fTB-perturbed-h-fieldeq} and \eqref{eq:fTB-trace-1st} allow for a full determination of the weak-field metric Eq.\eqref{eq:fTB-trace-reversed-def}. Observe that in vacuum, these equations give rise to gravitational waves which polarisation states have already been investigated in detail \cite{Farrugia:2018gyz}. 

\subsection{Solving the Field Equations}

In general, the solutions for $h_{\mu\nu}$ and $\mathring{R}$ can be obtained by making use of a Green's function $G(\textbf{x}, \textbf{x}^\prime)$, which result into 
\begin{align}
&\bar{h}_{\mu\nu} = \frac{4G}{c^4 f_T(0,0)} \int \frac{\Theta_{\mu\nu}\left(t-r/c,\textbf{x}^\prime\right)}{r} \, d^3 x^\prime, \\
&R = -\frac{\kappa^2}{3f_{BB}(0,0)}\int G_R(\textbf{x}, \textbf{x}^\prime) \Theta(\textbf{x}, \textbf{x}^\prime) \, d^4 x^\prime,
\end{align}
where $r = |\textbf{x}-\textbf{x}^\prime|$ and the Green's function $G_R$ defined as \cite{Berry:2011pb,Dass:2019kon}
\begin{equation}
G_R(\textbf{x}, \textbf{x}^\prime) = \frac{1}{(2\pi)^4}\int d^4 y \, \frac{e^{-i \textbf{y} \cdot \textbf{r}}}{\mu^2 - y^2}.
\end{equation}

Within the practical application of the weak-field approximation, it is sufficient to consider a slowly rotating source while keeping all terms up to the order of $c^{-3}$. Thus, the stress-energy tensorial components would be negligible within this context. In other words, the stress-energy tensor takes the form \cite{Ruggiero:2002hz,Mashhoon:1999nr}
\begin{equation}
\Theta_{\mu\nu} = \begin{pmatrix}
\rho c^2 & -\rho v_i c \\
-\rho v_i c & 0 
\end{pmatrix},
\end{equation}
where $\rho$ is the density of the source and $v_i$ is the velocity vector. Alternatively, the off-diagonal components can be simply expressed in terms of the mass current vector $j_i \coloneqq \rho v_i$. In this way, we therefore find that
\begin{align}
&\bar{h}_{00} = \frac{4G}{c^4 f_T(0,0)} \int \frac{\rho c^2}{r} \, d^3 x^\prime = \frac{4\Phi}{c^2 f_T(0,0)}, \label{metric00}\\
&\bar{h}_{0i} = -\frac{4G}{c^4 f_T(0,0)} \int \frac{j_i c}{r} \, d^3 x^\prime = -\frac{2A_i}{c^2 f_T(0,0)}\label{metric0i}\\
&\bar{h}_{ij} = 0 \\
&R = \frac{8 \pi G \mu^2}{c^2 f_T(0,0)}\int G_R(\textbf{x}, \textbf{x}^\prime) \rho\left(\textbf{x}^\prime\right) \, d^4 x^\prime.\label{metricij}
\end{align}
where $\Phi$ and $\textbf{A}$ are the scalar and vector potentials respectively. This yields the weak-field metric
\begin{align}
ds^2 &= -c^2 \left(1-\frac{2\alpha}{c^2 f_T(0,0)}\right)dt^2 - \frac{4}{c}(\textbf{A} \cdot d\textbf{x}) dt \nonumber \\
&+ \left(1+\frac{2\beta}{c^2 f_T(0,0)}\right)d\Sigma^2, \label{eq:fTB-weak-field-metric}\\
\alpha &\equiv \Phi - \frac{1}{2}f_{BB}(0,0) R c^2, \; \beta \equiv 2\Phi - \alpha,
\end{align}
where $d\Sigma^2 = dx^2 + dy^2 + dz^2$.

As the main aim of this work is to match with Gravity Probe B and Solar System observations, it is imperative to treat the source as a slowly rotating spherically symmetric static source having a constant mass $M$, radius $R_\text{S}$ and angular momentum $J$ with a constant density profile $\rho$ expressed, for simplicity, as 
\begin{equation}
\rho = \begin{cases}
\rho_0 = \dfrac{M}{\frac{4}{3}\pi {R_\text{E}}^3} & 0 \leq r \leq R_\text{S}, \\
0 & r > R_\text{S}.
\end{cases}
\end{equation} 
Under these assumptions, for distances sufficiently far away from the source (as the field is weak), the integrals can be solved through the Legendre polynomial expansion
\begin{equation}
\frac{1}{r} = \sum\limits_{l=0}^\infty \frac{L^{\prime l}}{\tilde{L}^{l+1}} \mathcal{P}_l \left(\cos \Theta\right),
\end{equation}
where $L = |\textbf{x}|$, $\tilde{L} = \left|\textbf{x}^\prime\right|$ and $\Theta$ is the angle between the two position vectors \textbf{x} and $\textbf{x}^\prime$. This yields the solutions \cite{Berry:2011pb}
\begin{align}
&\Phi = \frac{GM}{r}, \hspace{2cm}\textbf{A} = -\frac{G \textbf{J} \times \textbf{r}}{r^3 c}, \nonumber \\
&\mathring{R} = \frac{6\mu^2 \Phi e^{-\mu r}}{c^2 f_T(0,0)} \left[\frac{\mu R_\text{S}\cosh(\mu R_\text{S}) - \sinh(\mu R_\text{S})}{\mu^3 {R_\text{S}}^3}\right],
\end{align}
where $\textbf{J}$ represents the angular momentum vector. Therefore, the weak-field metric takes the simple form
{\small\begin{align}
&ds^2 = -c^2 \left\lbrace 1-\frac{2\Phi}{c^2 f_T(0,0)}\left[1+ e^{-\mu r}\eta(\mu R_\text{E}) \right]\right\rbrace dt^2 \nonumber \\ 
&- \frac{4}{c}(\textbf{A} \cdot d\textbf{x}) dt + \left\lbrace 1+\frac{2\Phi}{c^2 f_T(0,0)}\left[1- e^{-\mu r}\eta(\mu R_\text{E})\right]\right\rbrace d\Sigma^2, \label{eq:fTB-weak-field-metric-spherical-source}
\end{align}}
where we have defined the function
\begin{equation}
    \eta(x) \equiv \frac{x\cosh(x) - \sinh(x)}{x^3}.
\end{equation}

\subsection{Analogy with GEM\label{gem_sec}}

From the resulting weak-field metric, we can make a direct analogy with gravitoelectromagnetism (GEM) to generate the corresponding gravitoelectric and gravitomagnetic fields. Whilst these fields remain effectively unchanged in form, the Lorentz force is affected by the scalar $\mathring{R}$ mode similar to what is encountered in $f(\mathring{R})$ gravity. Following the steps dictated in Ref.\cite{Dass:2019kon}, the GEM equations and the Lorentz force equation are obtained as follows. Starting from the Lorenz gauge condition $\partial^\mu \bar{h}_{\mu \nu} = 0$, we obtain that 
\begin{equation}\label{eq:Lorentz_constraint}
\frac{1}{c}\frac{\partial \Phi}{\partial t} + \frac{1}{2} \nabla \cdot \textbf{A} = 0,
\end{equation}
with the remaining equations $\partial^\mu \bar{h}_{\mu i} = 0$ are of order $\mathcal{O}\left(c^{-4}\right)$ and therefore neglected. The gravitomagnetic field \textbf{B} and gravitoelectric field \textbf{E} are then defined as
\begin{align}
&B = \nabla \times \textbf{A}, & &E = -\nabla \phi - \frac{1}{2c}\frac{\partial \textbf{A}}{\partial t}.
\end{align}
It can then be easily shown that using Eqs.\eqref{eq:Lorentz_constraint} and \eqref{eq:fTB-perturbed-h-fieldeq}, the GEM equations result:
\begin{align}
&\nabla \cdot \textbf{E} = 4\pi G \rho, & &\nabla \cdot \textbf{B} = 0\\
&\nabla \times \textbf{E} = -\frac{1}{2c}\frac{\partial \textbf{B}}{\partial t}, & &\nabla \times \textbf{B} = \frac{2}{c}\frac{\partial \textbf{E}}{\partial t} + \frac{8\pi G}{c} \textbf{j}.
\end{align}

On the other hand, the Lorentz force for a test particle of mass $m$ can be obtained starting from its Lagrangian $\mathcal{L} = -mc \frac{ds}{dt}$, using the weak-field metric solution Eq.\eqref{eq:fTB-weak-field-metric} and expanding up to first-order of the potentials. This  yields
\begin{equation}
\mathcal{L} = -\frac{mc^2}{\gamma} + \frac{m\gamma }{f_T(0,0)}\left(\alpha + \beta \frac{v^2}{c^2}\right) - \frac{2m\gamma}{c}(\textbf{A} \cdot \textbf{v}), 
\end{equation} 
where $\gamma$ is the Lorentz factor and $\textbf{v} = \frac{d\textbf{x}}{dt}$ is the velocity vector. From the equations of motion $\frac{d}{dt}\left(\frac{\partial \mathcal{L}}{\partial \dot{\textbf{v}}}\right) = \frac{\partial \mathcal{L}}{\partial \textbf{x}}$, assuming that the vector potential \textbf{A} is stationary, it can be shown that up to first order in $v^2/c^2$, the force $\textbf{F} \equiv \frac{d\textbf{p}}{dt}$ where $\textbf{p} = m\gamma \textbf{v}$ is the relativistic momentum vector obeys
\begin{equation}\label{lorentz_force_like}
\textbf{F} = -\frac{m\textbf{E}}{f_T(0,0)} - \frac{2m}{f_T(0,0) c} (\textbf{v} \times \textbf{B}) + 3m \mu^2 c^2 \nabla \mathring{R}.
\end{equation}
Similar to Ref.\cite{Dass:2019kon}, one obtains the first two terms which are found in GR (except for a gravitational constant rescaling from $f_T(0,0)$) with a new contribution arising from the scalar mode. However, if the scalar mode is absent (i.e. $\mu^2 \to \infty$), the Lorentz force reduces to its GR form. 

\subsection{Comparison with a Spherically Symmetric Metric: The Schwarzchild Solution}

In the absence of rotation, the resulting weak-field metric Eq.\eqref{eq:fTB-weak-field-metric-spherical-source} cannot be directly correlated with the Schwarzchild solution due to the preferred choice of coordinates set by the Lorenz gauge. However, the metric can be transformed into a spherically symmetric form which can then be associated to such known solutions, and shall be notably important when discussing the geodetic effect. Here, we follow the procedure shown in Ref.\cite{Berry:2011pb}. The aim is to express the weak-field metric into the spherically symmetric form
\begin{equation}\label{eq:sphr-symm-metric}
    ds^2 = -c^2 A(\tilde{r}) dt^2 + B(\tilde{r}) d\tilde{r}^2 + \tilde{r}^2 d\Omega^2,
\end{equation}
with $A(\tilde{r})$ and $B(\tilde{r})$ representing some scalar functions and $d\Omega^2$ represents the polar symmetry. The necessary coordinate transformation is dictated by the condition
\begin{equation}
\tilde{r}^2 = \left(1+\frac{2\beta}{c^2 f_T(0,0)}\right)r^2 \implies \tilde{r} = \left(1+\frac{\beta}{c^2 f_T(0,0)}\right)r,
\end{equation}
where the last equality only holds for a weak-field. In particular, for a spherically symmetric static source, we have 
\begin{equation}
\tilde{r} = r+\frac{GM}{c^2 f_T(0,0)}\left[1- e^{-\mu r}\eta(\mu R_\text{S})\right].
\end{equation}
In this way, we obtain that up to first order in $M/\tilde{r}$
\begin{equation}
A(\tilde{r}) = 1-\frac{2GM}{c^2 f_T(0,0)\tilde{r}}\left[1+ e^{-\mu r}\eta(\mu R_\text{S}) \right].
\end{equation}
Observe that the exponential, similar to $f(R)$ gravity, retains the $r$ dependence. On the other hand, $B(\tilde{r})$ is found to be
\begin{align}
B(\tilde{r}) = &1 + \frac{2GM}{c^2 \tilde{r} f_T(0,0)}\left[1-e^{-\mu r}\eta(\mu R_\text{S})\right] \nonumber \\
&- \frac{2GM}{c^2 f_T(0,0)}\mu e^{-\mu r}\eta(\mu R_\text{S}).
\end{align}
Evidently, when $\mu \to \infty$ (i.e. in the limit of GR or when $f(T,B) \to f(T)$), the metric reduces to its Schwarzchild form.

\section{Perturbations on a Static Spherically Symmetric Metric: \texorpdfstring{$f(T)$}{fT}~Gravity \label{metric_per_sec}}

In the previous section, we have initially assumed that the gravitational field is weak, for which the relevant weak-field metric for an arbitrary $f(T,B)$ function was obtained. In what follows, a different approach is considered, particularly in the context of $f(T)$ gravity. Originally considered in Ref.\cite{DeBenedictis:2016aze} which was further pursued in Ref.\cite{Bahamonde:2019zea}, the idea is to assume a static spherically symmetric geometry arising due to a spherically symmetric static source of mass $M$. Then, one solves the field equations Eq.\eqref{eq:fTB-fieldeq-tetradform} to obtain the corresponding metric. Since no exact solutions have been obtained following this approach (although exact solutions can be found assuming, for instance, that the Lagrangian exhibits a Noether symmetry \cite{Paliathanasis:2014iva,Bahamonde:2019jkf}), a perturbative approach is employed where it is assumed that the $f(T)$ Lagrangian takes the form of
\begin{equation}
    f(T) = T + \epsilon \, F(T),
\end{equation}
where $\epsilon \ll 1$ represents a small, fiducial, order parameter, which will be omitted once the perturbations are solved. The role of the latter is to represent the small correction to the TEGR Lagrangian. In this way, the scalar functions $A(\tilde{r})$ and $B(\tilde{r})$ of the metric Eq.\eqref{eq:sphr-symm-metric} are expected to be in the form of a background solution plus a small correction. This will allow for corrections which were previously omitted in the weak-field regime. For simplicity, the source shall be assumed to be non-rotating as no perturbed solutions to the Kerr metric have been yet obtained.

In this formulation, the TEGR term gives rise to the exact Schwarzchild solution while the small correction sourced by $F(T)$ yields the first-order correction to the solution. Since the main interest lies in the Gravity Probe~B results, an alternative approach is to assume the gravitational field to be weak, meaning the metric can be approximated by a Minkowski spacetime background plus a small correction. Both approaches shall be presented and show that the same results are ultimately recovered, whilst offering a more detailed analysis on the affect of the $f(T)$ Lagrangian on the geodetic effect.

\subsection{Perturbations on the Schwarzchild Solution}

The Schwarzchild correction can be obtained by taking the scalar potentials to be expressed as
\begin{align}
    A(\tilde{r}) &= 1 - \frac{2GM}{c^2 \tilde{r}} + \epsilon \, \mathcal{A}(\tilde{r}), \\
    B(\tilde{r}) &= \left(1-\frac{2GM}{c^2 \tilde{r}}\right)^{-1} + \epsilon \, \mathcal{B}(\tilde{r}),
\end{align}
for some functions $\mathcal{A}$ and $\mathcal{B}$. To solve for the corrections, the field equations are perturbed up to first order in $\epsilon$. For simplicity, the power-law ansatz $F(T) = \alpha T^p$ for some constant $\alpha$ and $p$ is considered. Furthermore, unless otherwise stated, $\frac{GM}{c^2} \to M$.

\begin{widetext}
The solutions for the scalar functions can be obtained from the differential equations
\begin{align}
    &\mathcal{B}_x = \frac{2 \left(1-2 x^2\right) \mathcal{B}}{x-x^3}+\frac{2 \alpha  (-1)^p (p-1) M^{2-2 p} (x-1)^{2 p-4} x^{6-5 p} \left((p-1) x^2+2 p x+5 p+x\right) \left(x^2-1\right)^{2 p}}{(x+1)^3}, \\
    &\mathcal{A}_{\tilde{r}}-\frac{2 M \mathcal{A}}{\tilde{r}(\tilde{r}-2 M)} = \left(1-\frac{2 M}{\tilde{r}}\right)\mathcal{B}+\alpha  2^{3 p-2} (1-p) \tilde{r}^{1-3p}\left[\tilde{r} + \sqrt{\frac{\tilde{r}}{\tilde{r}-2 M}}\left(M-\tilde{r}\right)\right]^p,
\end{align}
where $x \coloneqq \left(1-\frac{2M}{\tilde{r}}\right)^{-\frac{1}{2}}$. Although a general solution is not recovered in general, some special cases are considered. For $p = 2$, the scalar functions are given to be \cite{DeBenedictis:2016aze,Bahamonde:2019zea} 
{\small\begin{align}
    A(\tilde{r}) &= 1 -\frac{2 M}{\tilde{r}} + \alpha \left(\frac{32 }{3 M^2}\left[-1+\left(1-\frac{2 M}{\tilde{r}}\right)^{3/2}\right]-\frac{(\tilde{r}-3 M)}{M^2 \tilde{r}}\ln \left(1-\frac{2 M}{\tilde{r}}\right) -\frac{2  M}{\tilde{r}^3}+\frac{30 }{M \tilde{r}}-\frac{12 }{\tilde{r}^2}\right), \label{eq:fT-weakSch-A-p2}\\
    B(\tilde{r}) &= \frac{1}{1-\frac{2 M}{\tilde{r}}}+\frac{\alpha}{Mr(1-\frac{2 M}{\tilde{r}})^2} \left[-\frac{2 \left(75 M^2-69 M \tilde{r}+16 \tilde{r}^2\right)}{3 \tilde{r}^2} +\frac{16 (3 M-\tilde{r}) (M-2 \tilde{r})}{3 \tilde{r}^2}\left(1-\frac{2M}{\tilde{r}}\right)^{1/2}-\ln \left(1-\frac{2 M}{\tilde{r}}\right)\right]. \label{eq:fT-weakSch-B-p2}
\end{align}}
\end{widetext}
The solutions are well behaved in the sense that in the absence of a source, the solutions reduce to Minkowski space as expected, i.e. when $M \to 0$, \mbox{$A(\tilde{r}), B(\tilde{r}) \to 1$}. If the gravitational field is weak, the solutions follow the order expansion 
\begin{align} 
    A(\tilde{r}) &= 1-\frac{2 M}{\tilde{r}}-\frac{16 \alpha  M^3}{5 \tilde{r}^5} + \mathcal{O}\left(\frac{M^6}{\tilde{r}^6}\right), \\
    B(\tilde{r}) &=1+\frac{2 M}{\tilde{r}} +\frac{4 M^2}{\tilde{r}^2} +\frac{8 M^3}{\tilde{r}^3} +\frac{16 \alpha  M^3 }{\tilde{r}^5} + \mathcal{O}\left(\frac{M^6}{\tilde{r}^6}\right).
\end{align}
which agrees with the weak-field metric Eq.\eqref{eq:fTB-weak-field-metric-spherical-source} in the limit of $f(T,B) \to f(T) = T + \alpha T^p$ up to first order in $M/\tilde{r}$. Observe that the $\alpha$ contributions do not appear in the latter as it is a higher-order contribution. Solutions for other values of $p$ are considered in Ref.\cite{Bahamonde:2019zea}. For the purpose of the analysis which follows, the solution for $p = 3$ is given and is listed in Appendix \ref{app:p-3_sol}. 

\subsection{An Alternative Approach for a Weak-Field Limit}

If the field is assumed to be weak, the scalar functions can be expanded around a Minkowski background according to
\begin{align}
    A(\tilde{r}) &= 1 + \epsilon A_1(\tilde{r}) + \epsilon^2 A_2(\tilde{r}) + \epsilon^3 A_3(\tilde{r}) + \dots{}, \\
    B(\tilde{r}) &= 1 + \epsilon B_1(\tilde{r}) + \epsilon^2 B_2(\tilde{r}) + \epsilon^3 B_3(\tilde{r}) + \dots{}.
\end{align}
Once again, assuming that the Lagrangian $f(T)$ is Taylor expandable about $T = 0$, and solving the field equations order by order yields 
\begin{align}
    A(r) &= 1+\epsilon  \left(c_2-\frac{c_1}{\tilde{r}}\right)+\epsilon ^2 \left(c_4-\frac{c_1 c_2+c_3}{\tilde{r}}\right) \nonumber \\
    &+\epsilon ^3 \left(c_6 - \frac{c_1^3 f_{TT}(0)}{5 \tilde{r}^5 f_T(0)}-\frac{c_2 c_3+c_1 c_4+c_5}{\tilde{r}} \right), \\
    B(r) &= 1+\frac{c_1 \epsilon }{\tilde{r}}+\epsilon ^2 \left(\frac{c_1^2}{\tilde{r}^2}+\frac{c_3}{\tilde{r}}\right) \nonumber \\
    &+\epsilon ^3 \left(\frac{c_5}{\tilde{r}}+\frac{c_1^3 f_{TT}(0)}{\tilde{r}^5 f_T(0)}+\frac{c_1^3}{\tilde{r}^3}+\frac{2 c_3 c_1}{\tilde{r}^2}\right),
\end{align}
where $c_{1,\text{\dots{}},6}$ are integration constants. To determine these constants, we impose the following constraints. As $\tilde{r} \to \infty$ (i.e. far away from the source), the metric must reduce to Minkowski spacetime and thus sets $c_{2,4,6} = 0$. On the other hand, according to the solution obtained in \S.\ref{weak_field_approx}, namely Eq.\eqref{eq:fTB-weak-field-metric-spherical-source}, we find that $c_1 = \frac{2M}{f_T(0)}$ (alternatively, it can be reasoned that in the limit of TEGR, the metric must reduce to the Schwarzchild metric). Finally, the constants $c_{3,5}$ have to be zero otherwise the solution does not reduce to its TEGR limit for $f(T) = T$. Therefore, the final solution is
\begin{align}
    A(r) &= 1-\frac{2 M}{\tilde{r} f_T(0)}-\frac{8 M^3 f_{TT}(0)}{5 \tilde{r}^5 {f_T(0)}^4}, \\
    B(r) &= 1+\frac{2 M}{\tilde{r} f_T(0)}+\frac{4 M^2}{\tilde{r}^2 {f_T(0)}^2}+\frac{8 M^3}{\tilde{r}^3 {f_T(0)}^3}\nonumber \\
    &+\frac{16 M^4}{\tilde{r}^4 {f_T(0)}^4}+\frac{8 M^3 f_{TT}(0)}{\tilde{r}^5 {f_T(0)}^4}.
\end{align}
Taking $f(T) = T + \alpha T^2$ recovers the previously obtained weak-field limit solution as expected. Observe that for $f(T) = T + \alpha T^n$, $n > 2$ ($n$ integer) does not reveal any contributions at this order meaning their effects are even smaller. On the other hand, this approach is not applicable for functions which are not expandable about $T = 0$, for instance $f(T) = T + \alpha T^n$, $n < 0$ and even some cosmologically viable ones such as the Linder model $f(T) = T + \alpha T_0 \left(1- e^{-p \sqrt{T/T_0}}\right)$ for some constant $p$. However, there exist cosmological model Lagrangians which may be further investigated for such weak-field observational tests, such as $f(T) = T + \alpha T_0 (1- e^{-p T/T_0})$ and $f(T) = T + \alpha T^n \tanh(T/T_0)$ for appropriate values of $p$ and $n$ \cite{Nesseris:2013jea}. 

Observe that the result is in agreement with the parameterised post-Newtonian (PPN) approximation investigated in Ref.\cite{Ualikhanova:2019ygl} since there is no deviation up to second order expansion. The first modification appears at third order when the $f_{TT}$ term contributes to the behaviour.

\section{Observational Constraints\label{ob_const_sec}}

As we have now closely discussed the theoretical foundations to obtain the necessary metrics, in what follows, we apply those results to observations obtained by Gravity Probe~B and from classical Solar System test observations. In particular, we shall focus on the geodetic effect (de Sitter precession), the Lense-Thirring effect, Shapiro time delay, light bending and perihelion precession.

\subsection{Geodetic Effect}

The geodetic effect describes the effect of a precessing gyroscope due to its orbit around a massive central body. Here, we obtain the precession rate following Rindler's approach \cite{Rindler:2006km}. Starting from a spherically symmetric metric, we consider the system to be rotating at an angular frequency $\omega$, i.e.
\begin{equation}\label{coor_trans_geodetic}
\phi = \phi^\prime - \omega t.
\end{equation}
By assuming the gyroscope to lie in a circular polar orbit (at an angle $\theta = \frac{\pi}{2}$) allows us to rewrite the metric in a canonical form
\begin{equation}
ds^2 = -e^{2\Psi} \left(dt - e^{-2\Psi} \omega \tilde{r}^2 d\phi^\prime\right)^2 + A r^2 e^{-2\Psi} d{\phi^\prime}^2,
\end{equation}
where $e^{2\Psi} \equiv A - r^2 \omega^2$. As shown in Ref.\cite{Rindler:2006km}, the angular frequency of the gyroscope is given to be
\begin{align}
\Omega &= \frac{e^\Psi}{2\sqrt{2}}\left[k^{ik} k^{jl} (\omega_i,j - \omega_j,i)(\omega_k,l - \omega_l,k)\right]^{1/2},
\end{align}
where $k^{ij}$ is the spatial 3-metric and $\omega_i \equiv e^{-2\Psi} \omega r^2 \delta^3_i$, which simplifies to
\begin{equation}
\Omega = \frac{\omega}{\sqrt{AB}}.
\end{equation}
The angle after one full revolution is then given to be $\alpha^\prime = \Omega \Delta \tau$, where $\Delta\tau$ represents the proper time of the gyroscope, which can be obtained directly from the metric
\begin{align}
&d\tau^2 = A dt^2 - \tilde{r}^2 d\phi^2 = Adt^2 - \tilde{r}^2 \omega^2 dt^2 \nonumber \\
&\implies \Delta\tau = \sqrt{A - \tilde{r}^2 \omega^2} \Delta t.
\end{align}
Thus, the precession over one orbit is $\alpha = 2\pi - \alpha^\prime$, which implies that  the precession rate per year is given to be
\begin{equation}\label{geodetic_precess}
\Omega_{\text{dS}} = \sqrt{\frac{A_{\tilde{r}}}{2\tilde{r}}}\left[1-\sqrt{\frac{1}{B}\left(1-\frac{\tilde{r}A_{\tilde{r}}}{2A}\right)}\right].
\end{equation}

\subsection{Lense-Thirring Precession}

It is well known that the Lense-Thirring precession in GR can be simply derived by assuming a freely falling gyroscope initially at rest with an angular spin vector $S^\mu$. Taking $u^\mu$ to represent the gyroscope's rest frame velocity, we have that $S^\mu u_\mu = 0$. Then, the Lense-Thirring precession rate would be then obtained using the geodesic equations
\begin{equation}
\frac{dS^\mu}{d\tau} + \mathring{\Gamma}^\mu_{\sigma\rho} S^\sigma u^\rho = 0.
\end{equation}
In the context of teleparallel gravity, the gyroscope moves according to force-like equations
\begin{equation}
\frac{dS^\mu}{d\tau} + \udt{\Gamma}{\mu}{\sigma\rho} S^\sigma u^\rho = \udt{K}{\mu}{\sigma\rho} S^\sigma u^\rho.
\end{equation}
Despite this apparent difference, the above is mathematically equivalent to the geodesic equation due to the fact that $\udt{K}{\sigma}{\mu\nu} = \mathring{\Gamma}^{\sigma}_{\mu\nu} - \Gamma^{\sigma}_{\mu\nu}$. Nonetheless, the force-like equations offer a different interpretation as discussed, for instance, in Refs.\cite{Aldrovandi:2006cy,Aldrovandi:2008xv}, as the teleparallel force equations allow for a separation between gravitation and inertia which has important implications on the weak equivalence principle (WEP), which lies beyond the scope of this manuscript (see, for instance, Ref.\cite{Aldrovandi:2003pa} for further discussions on the topic). Within the assumption that the WEP holds, one can follow the same steps encountered in GR. Alternatively, one can work out directly using the torsion and contorsion tensors to obtain the same result.

If the field is weak, the field equations reduce to 
\begin{equation}\label{lense_thirring_GPB}
\frac{dS_i}{d\tau} = \epsilon_{ikl} \Omega^k S^l,
\end{equation}
where $\Omega^k \equiv -\frac{1}{2}\epsilon^{kmn}\partial_m h_{0n}$ defines the angular velocity precession vector of the gyroscope. Following the results obtained in the $f(T,B)$ weak-field solution Eq.\eqref{eq:fTB-weak-field-metric-spherical-source}, we find that the Lense-Thirring precession rate $\Omega_\text{LT}$ remains unaffected except for a Newtonian rescaling, which is expected as the electromagnetic field is identical to that found in GR. However, this result is only valid within the context of weak-fields and thus remains to be investigated in the case of strong gravitational fields. 

\subsection{Shapiro Time Delay}

The effect of Shapiro time delay \cite{Shapiro:1964uw} can be derived following the steps listed in Ref.\cite{Weinberg:1972kfs}. Here, we focus on deriving the $\alpha$ dependent correction for the $f(T)$ power-law model. For the given spherically symmetric metric Eq.\eqref{eq:sphr-symm-metric}, the time delay of a radio signal as it travels from the Earth to Mercury and back, as the signal passes through the closest point of approach $R \simeq R_{\odot}$ to the Sun is 
\begin{align}
    \Delta t &= 2 \Big[t \left(r_{\oplus}, {R_{\odot}} \right) + t \left(r_{\mercury}, {R_{\odot}} \right) \nonumber\\
    &- \sqrt{{r_{\oplus}}^2-{{R_{\odot}}^2}} - \sqrt{{r_{\mercury}}^2-{{R_{\odot}}^2}} \Big].
\end{align}
where $r_{\oplus}$ and $_{\mercury}$ represent the Earth and Mercury orbital radii respectively, and $t(\tilde{r},R)$ is defined as 
\begin{equation}
    t(\tilde{r},R) = \int\limits_{R}^{\tilde{r}} \frac{d\bar{r}}{\sqrt{\left(1-\frac{R^2 A(R)}{\bar{r}^2 A(\bar{r})}\right)\frac{A(\bar{r})}{B(\bar{r})}}}.
\end{equation}
Using the fact that generally, the orbital radii satisfy the condition $\tilde{r} \gg R$, together with the weak-field metric solutions Eqs.\eqref{eq:fT-weakSch-A-p2},\eqref{eq:fT-weakSch-B-p2},\eqref{eq:fT-weakSch-A-p3},\eqref{eq:fT-weakSch-A-p3}, we find that the $\alpha$ contribution takes the following forms
\begin{align}\label{eq:shapiro_delay_alpha}
    &t_\alpha(\tilde{r},R) \approx \nonumber\\
    &\begin{cases}
    \frac{4 \alpha  M^3}{3} \left(\frac{32}{3\tilde{r}^4}-\frac{2}{\tilde{r}R^3} -\frac{1}{\tilde{r}^3 R}-\frac{2}{R^4}\right), & p = 2, \\
    \frac{\alpha  M^5}{630} \left(-\frac{4608}{R^8}+\frac{560}{\tilde{r}R} +\frac{280}{\tilde{r}^3 R^5} + \frac{210}{\tilde{r}^5 R^3}+\frac{175}{\tilde{r}^7 R}+\frac{700}{\tilde{r}^8}\right), & p = 3.
    \end{cases}
\end{align}



\subsection{Light Bending}

The total deflection angle of light is derived following the method used in Ref.\cite{Farrugia:2016xcw}. A photon is assumed to be emitted from some far away source at an angle $\phi = -\pi/2$. It travels and reaches a point of closest approach $\tilde{r} = \tilde{r}_\star$ with respect to some spherical massive source at $\phi = 0$, and then continues to travel away from the source approaching an angle of $\phi = \pi/2$. To account for the deflection angle due to the gravitational attraction of the source, we start from the spherically symmetric metric Eq.\eqref{eq:sphr-symm-metric} within the equatorial plane $\theta = \pi/2$ to obtain that the path of the photon obeys the second order differential equation
\begin{align}
0 &= \dfrac{d^2 u}{d\phi^2} + \dfrac{1}{2AB} \dfrac{d}{du} \left[u^2 A\right] \nonumber \\
&+ \dfrac{1}{2AB} \dfrac{d}{du} \left[\ln(AB) \right] \left({u_R}^2- u^2 A\right), 
\end{align}
where $u = 1/\tilde{r}$ and $u_R = 1/R$ is the inverse impact parameter, with boundary conditions $u(0) = u_\star = 1/\tilde{r}_\star$ and $u(\pm \pi/2) = 0$. Since the differential equation cannot be solved in general, even for weak-field sources, the perturbative iterative method considered in Ref.\cite{Bodenner2003} is applied.

The approach aims to obtain a perturbative solution by taking the mass $M$ as the perturbation parameter, i.e. we let
\begin{equation}
    u = u_0 + u_1 + u_2 + \text{\dots{}}
\end{equation}
where $u_i$ represents the solution up to $\mathcal{O}\left(M^i\right)$. As an illustrative example, we derive the perturbative solution for the power-law model with $p = 2$. In this case, the differential equation up to $\mathcal{O}(M^3)$ is
\begin{equation}
    0 = u^{\prime\prime} +u -3 M u^2 + \alpha M^3 u^4 \left(\frac{32}{R^2}-56 u^2\right),
\end{equation}
where we have denoted primes to represent derivatives with respect to $\phi$. This yields the following ordered system of differential equations
\begin{align*}
    {u_0}^{\prime\prime} &= -u_0, \\
    {u_1}^{\prime\prime} &= - u_1  + 3 M {u_0}^2 \\
    {u_2}^{\prime\prime} &= - u_2 +6 M u_0 u_1 \\
    {u_3}^{\prime\prime} &= -u_3 +3 M \left({u_1}^2+2u_0 u_2\right) - \alpha M^3 {u_0}^4 \left(\frac{32}{R^2}-56 {u_0}^2\right),
\end{align*}
which yields the final expression for $u$ to be
\begin{align}
    u &= -\frac{\alpha  M^3 (55 \cos (2 \phi )+8 \cos (4 \phi )+\cos (6 \phi )-80)}{20 R^6}\nonumber\\
    &-\frac{M^3 (60 \phi  \sin (2 \phi )+125 \cos (2 \phi )+\cos (4 \phi )-312)}{16 R^4}\nonumber\\
    &+\frac{3 M^2 (20 \phi  \sin (\phi )+22 \cos (\phi )+\cos (3 \phi ))}{16 R^3}\nonumber\\
    &-\frac{M (\cos (2 \phi )-3)}{2 R^2}+\frac{\cos (\phi )}{R}
\end{align}
Once the solution for $u$ is obtained, following Rindler and Ishak's approach \citep{Rindler:2007zz}, it can be shown that the total deflection angle is given to be
\begin{equation}
\epsilon \approx \left.\frac{2u}{|\frac{du}{d\phi}| \sqrt{B}}\right|_{\phi = \frac{\pi}{2}}.
\label{Psi-Approx}
\end{equation}
which, for the quadratic $f(T)$ Lagrangian yields the following solution
\begin{equation}\label{eq:deflection_alpha_2}
    \epsilon = \frac{4 M}{R}+\frac{15 \pi  M^2}{4 R^2}+\frac{189 M^3}{4 R^3}+\frac{256 \alpha  M^3}{15 R^5} + \mathcal{O}(M^4).
\end{equation}
A similar analysis for the cubic $f(T)$ Lagrangian reveals that the total deflection angle is
\begin{align}
    \epsilon &= \frac{4 M}{R}+\frac{15 \pi  M^2}{4 R^2}+\frac{189 M^3}{4 R^3}+\frac{4335 \pi  M^4}{64 R^4} \nonumber \\
    &+\frac{7155 M^5}{8 R^5}+\frac{225 \pi ^2 M^5}{32 R^5}-\frac{4096 \alpha  M^5}{315 R^9} + \mathcal{O}(M^6). \label{eq:deflection_alpha_3}
\end{align}
Observe that in both cases, the GR second order mass correction found in Refs.\cite{Bhattacharya:2009rv,Ishak:2010zh,Bodenner2003} is recovered. In general, for a Taylor expandable $f(T)$ model within the regime of weak gravitational fields, the first deviation from GR appears at $\mathcal{O}(M^3)$, having the form
\begin{equation}
    \Delta\epsilon = \frac{128 \alpha f_{TT}(0) M^3}{15 f_T(0)^4 R^5} + \mathcal{O}(M^4).
\end{equation}
Naturally, the quadratic weak-field result is recovered while for the cubic case requires the higher order contributions.

\subsection{Perihelion Precession}

The effect of $\alpha$ for the power-law ansatz Lagrangian on perihelion precession has been investigated in great detail in Refs.\cite{DeBenedictis:2016aze,Bahamonde:2019zea}. Here, we shall only quote the results:\footnote{A factor of 2 has been included to correctly match with the definition of $\alpha$ used in those works.}
\begin{align}
    &p = 2 & &\Delta\phi = \frac{16\pi\alpha M^2}{{r_\text{c}}^4} \label{eq:perihelion_alpha_2}\\
    &p = 3 & &\Delta\phi = -\frac{96\pi\alpha M^4}{{r_\text{c}}^8}. \label{eq:perihelion_alpha_3}
\end{align}
where $r_\text{c}$ represents the circular radius of the orbit. For the $n = 2$ case, the detailed analysis in Ref.\cite{DeBenedictis:2016aze} leads to a bound of $\alpha \lesssim \SI{E20}{\kilo\metre^2}$. 

\section{Numerical Results \label{obs_fitting}}

In this section, we make use of the weak-field solutions listed in \S.\ref{weak_field_approx} and \ref{metric_per_sec} against observations in order to constrain the Lagrangian free model parameters depending on the model considered. It is important, however, to comment on the results for an arbitrary $f(T,B)$ model for the case when $\mu \not\to \infty$. 

Although the weak-field metric has been obtained in its spherically symmetric form, the scalar functions $A(\tilde{r})$ and $B(\tilde{r})$ are not truly expressed in terms of $\tilde{r}$ since the relation between $r$ and $\tilde{r}$ is not invertible. This leaves two unknown parameters, the isotropic radial coordinate $r$ and $\mu$. However, $r$ is not measured and hence one must instead opt to impose specific values of $\mu$ to determine whether the results would then be consistent. Since the goal is to constrain the Lagrangian parameters (and hence constrain $\mu$ through observations), this option is not investigated in detail. Nonetheless, if $\mu$ is sufficiently large, the contributions would be small enough that deviations from observations (and hence form GR) are expected to be effectively negligible.

On the other hand, a more thorough investigation can be inferred in the case of $f(T)$ gravity using the results obtained in \S.\ref{metric_per_sec}. In particular, we shall make use of the results for the two power-law ansatz values considered, namely $p=2$ and 3, which eventually lead to observation constraints on the constant $\alpha$. 

\subsection{Geodetic Effect}

In April 2004, Gravity Probe B was launched starting its year and a half flight mission, with the purpose of accurately measuring the geodetic and the frame dragging precession rates while in orbit about the earth. A geodetic precession rate of $-6601.8\pm18.3$ mas/yr was measured while in a polar orbit at around \SI{642}{\kilo\metre} \cite{Gravity_Probe_B_1}.

Through the use of Eq.\eqref{geodetic_precess}, the $\alpha$ constraints are obtained as listed in Table~\ref{tab:Constraint_deltaOmega_equations}. The table also illustrates the $\alpha$ constraint which has to be obeyed for the weak-field approximation to hold (which is a direct consequence of the assumption that the perturbation $F(T) \ll T$).

\begin{table}[t!]
\centering
\begin{tabularx}{\columnwidth}{ccX}
\hline
$p$&$\alpha \ll $&\hspace{3cm}$\Omega_{\text{dS}}$\\
\hline
\\
$2$ & $\frac{5 \tilde{r}^2}{4 \mathcal{M}^2}$ & $\frac{3 c \mathcal{M}^{3/2}}{2 \tilde{r}} \bigg(1+\frac{3 \mathcal{M}}{4}+\frac{9 \mathcal{M}^2}{8}+\frac{135 \mathcal{M}^3}{64}+\frac{567 \mathcal{M}^4}{128}+\frac{12 \alpha  \mathcal{M}^2}{\tilde{r}^2}\bigg)$ \\\\
$3$ & $\frac{9 \tilde{r}^4}{8 \mathcal{M}^4}$ & $\frac{3 c \mathcal{M}^{3/2}}{2 \tilde{r}}\bigg(1+\frac{3 \mathcal{M}}{4}+\frac{9 \mathcal{M}^2}{8}+\frac{135 \mathcal{M}^3}{64}+\frac{567 \mathcal{M}^4}{128}+\frac{12 \alpha  \mathcal{M}^2}{r^2}+\frac{5103 \mathcal{M}^5}{512}+\frac{24057 \mathcal{M}^6}{1024}+\frac{938223 \mathcal{M}^7}{16384}+\frac{4691115 \mathcal{M}^8}{32768}-\frac{72 \alpha  \mathcal{M}^4}{\tilde{r}^4}\bigg)$\\\\\hline
\end{tabularx}
\caption{Illustration of the $\alpha$ weak-field constraint depending on the index $p$ for the $f(T)$ model $f(T) = T + \alpha T^p$ alongside the resulting Geodetic precession expressions based on the scalar functions $A(\tilde{r})$ and $B(\tilde{r})$ derived in Refs.\cite{DeBenedictis:2016aze,Bahamonde:2019zea}. Here, we have defined the parameter $\mathcal{M} \coloneqq \frac{GM}{c^2 \tilde{r}}$.}
\label{tab:Constraint_deltaOmega_equations}
\end{table}

Based on the expressions listed in Table \ref{tab:Constraint_deltaOmega_equations}, the corresponding numerical constraints are then obtained as shown in Table \ref{tab:Values}. Evidently, the constraints obtained from observations are well within the expected bounds of the weak-field condition which supports the consistency of the weak-field approach.

\begin{table}[t!]
\centering
\begin{tabularx}{\columnwidth}{p{1cm}p{2cm}X}
\hline
$p$ & $\alpha \ll $ & $\alpha_{\text{GPB}}/\si{\kilo\metre}^p$\\ \hline \\
$2$ & $\sim 10^{32}$ & $-7.5476 \times 10^{28} < \alpha < 3.8438 \times 10^{28}$\\ \\
$3$ & $\sim 10^{64}$ & $-2.3716 \times 10^{60} < \alpha < 4.6568 \times 10^{60}$\\ \\
\hline
\end{tabularx}
\caption{The numerical constraints for the constant $\alpha$ for the power-law model $f(T) = T + \alpha T^p$ where $p = 2, \, 3$ are set based on the Gravity Probe~B observations. Furthermore, an order of magnitude estimate where the weak-field approximation is valid has also been given.}
\label{tab:Values}
\end{table}

\subsection{Classical Solar System Constraints}

For Shapiro time delay and light deflection, the PPN formulation together with observations from the Cassini spacecraft pose a viable opportunity to obtain constraints. As illustrated, for instance in Ref.\cite{Weinberg:1972kfs,Misner:1974qy}, the $\gamma$ PPN parameter appears in the former tests as follows. For Shapiro time delay, the deviation from GR amounts to
\begin{equation}
\Delta t^{\text{PPN}} \simeq 4M \left(\dfrac{\gamma-1}{2}\right) \ln \left(\frac{4 r_{\oplus} r_{\mercury}}{{R_{\odot}}^2} \right).
\end{equation}
while for light bending, the total deflection angle is 
\begin{equation}
    \epsilon = \left(\frac{1+\gamma}{2}\right) \frac{4M}{R}.
\end{equation}
Using Cassini's experimental value of $\gamma -1 = \left(2.1 \pm 2.3\right) \times 10^5$ and the expressions Eqs.\eqref{eq:shapiro_delay_alpha}, \eqref{eq:deflection_alpha_2}, \eqref{eq:deflection_alpha_3}, \eqref{eq:perihelion_alpha_2} and \eqref{eq:perihelion_alpha_3}, $\alpha$ constraints are obtained as summarised in Table \ref{tab:solar_alpha_constraints}. In the case of perihelion precession, the $\alpha$ constraints are based on the observed precession rate of Mercury as investigated in Ref.\cite{DeBenedictis:2016aze}. 

\begin{table}[!htb]
    \centering
    \begin{tabularx}{0.8\columnwidth}{>{\centering\arraybackslash}Xc}
    \hline
    $p$ & $\alpha/\si{\kilo\metre}^p$ \\ \hline \\ 
    \multicolumn{2}{l}{\textbf{Shapiro Delay}} \\ \hline \\
    $2$ & $-8.26031\times 10^{16} < \alpha < 1.81727\times 10^{18}$ \\
    $3$ & $-3.78768\times 10^{44} < \alpha < 1.72167\times 10^{43}$ \\ \\ 
    \multicolumn{2}{l}{\textbf{Light Bending}} \\ \hline \\
    $2$ & $-1.82378\times 10^{17} < \alpha < 3.95829\times 10^{17}$ \\
    $3$ & $-5.57249\times 10^{43} < \alpha < 2.56754\times 10^{43}$ \\ \\ 
    \multicolumn{2}{l}{\textbf{Perihelion Precession}} \\ \hline \\
    2 & $\alpha < 2.23602\times 10^{20}$ \\
    3 & $\alpha > -8.18149\times 10^{49}$ \\ \\
    \hline
    \end{tabularx}
    \caption{A summary of the parameter constraints obtained for the coupling parameter, $\alpha$, for the power-law model $f(T) = T + \alpha T^p$ where $p = 2, \, 3$ using observations from perihelion precession and the Cassini spacecraft. Here, $M = M_{\odot} = \SI{1.47}{\kilo\metre}$, $R \simeq R_{\odot} = \SI{6.9551E5}{\kilo\metre}$, $r_{\oplus} = \SI{1.4710E11}{\kilo\metre}$, $r_{\mercury} = \SI{4.6001E7}{\kilo\metre}$ and $r_c = \SI{5.55E7}{\kilo\metre}$.}
    \label{tab:solar_alpha_constraints}
\end{table}

\section{Conclusion \label{sec:conclusion}}

The main result of this work is that both the classical solar systems and the gravitomagnetic constrains from Gravity Probe B result in a constant on the coupling parameter to $|\alpha| \lesssim \SI{E16}{\kilo\metre^2}$ for $p = 2$ and $|\alpha| \lesssim \SI{E43}{\kilo\metre^3}$ for $p=3$. This forms one of the strongest constraints on this parameter (to the best of our knowledge).

Gravitomagnetic effects are imperative for understanding the weak field limit of modified gravity in the context of rotation. In this work, we have explored these related effects in the TG framework. TG offers a novel possibility of constructing gravitational theories in which the background manifold is torsionful rather than curvatureful. While this is dynamically equivalent to GR in the TEGR limit, modifications of the TEGR action produce theories which may be distinct from what can be constructed in regular curvature-based theories of gravity. This allows for the possibility of totally new models of gravity that may have important consequences for meeting the observational challenges of the coming years.

The main crux of the weak field analysis stems from the analysis in \S.\ref{weak_field_approx} where we take an order by order expansion of a general $f(T,B)$ gravity Lagrangian. In Eq.(\ref{eq:fTB-trace-1st}), this is found to potentially behave as a massive theory with a mass that is mainly dependent on whether a boundary term contribution is present or not. This approximation is then set into the field equations with a slowly rotating source to find metric solutions in Eqs.(\ref{metric00}--\ref{metricij}). In \S.\ref{gem_sec} we go into the details of how this analogy tallies with the well-known GEM effects to produce a Lorentz force-like effect in Eq.(\ref{lorentz_force_like}). Finally, we compare this with the Schwarzschild solution to determine the relation to the effective mass of the general $f(T,B)$ model.

Limiting ourselves to $f(T)$ gravity, we explore the possibility of perturbative solutions in \S.\ref{metric_per_sec} where exact solutions are found up to perturbative order in the spherically symmetric setting. These were also investigated in the literature \cite{DeBenedictis:2016aze,Bahamonde:2019zea} and remain an interesting avenue of research in the TG context. In this part of the work, we investigate two possible routes to the perturbative analysis which both agree in their PPN limit.

The traditional gravitomagnetic effects of the geodetic and Lense-Thirring phenomena are determined in \S.\ref{ob_const_sec}. The Geodetic effect naturally emerges for a static system with a rotating observer. This is achieved by a coordinate transformation, as prescribed in Eq.(\ref{coor_trans_geodetic}). This eventually produces Eq.(\ref{geodetic_precess}) which is our result for the geodetic precession rate and the main result of that subsection. The Lense-Thirring effect is then determined for this TG case where the main result result is shown in Eq.(\ref{lense_thirring_GPB}) which is comparable to the Gravity Probe B mission result. In fact, in \S.\ref{obs_fitting} we use the results of this mission to constrain our parameters for the various potential models under investigation.

Gravitomagnetic effects have the potential to have an important impact on understanding which modified theories of gravity are viable and may play an important role in the coming years for developing realistic modified theories of gravity.





%

\begin{widetext}
\appendix
\section{Solution for \texorpdfstring{$p = 3$}{p3} \label{app:p-3_sol}}

The perturbed spherically symmetric metric for the model ansatz $f(T) = T + \alpha T^3$ is given to be
\begin{align}
    A(\tilde{r}) &= 1-\frac{2 M}{\tilde{r}}+\frac{\alpha }{M^4}\left[\frac{2\left(7M -3 \tilde{r} \right)}{\tilde{r}} \ln \left(1-\frac{2 M}{\tilde{r}}\right)-\frac{8 M^5}{\tilde{r}^5}-\frac{136 M^4}{9 \tilde{r}^4}+\frac{12 M^3}{\tilde{r}^3} +\frac{16 M^2}{\tilde{r}^2}+\frac{2476 M}{135 \tilde{r}}-\frac{4096}{315} \right.\nonumber \\
    & + \sqrt{\frac{\tilde{r}}{\tilde{r}-2 M}}\left(-\frac{128 M^6}{27 \tilde{r}^6}+\frac{128 M^5}{27 \tilde{r}^5}+\frac{33536 M^4}{945 \tilde{r}^4} \left. +\frac{8192 M^3}{945 \tilde{r}^3}+\frac{22528 M^2}{945 \tilde{r}^2}-\frac{8192 M}{189 \tilde{r}}+\frac{4096}{315}\right) \right], \\
    B(\tilde{r}) &= \frac{1}{1-\frac{2 M}{\tilde{r}}}+ \frac{\alpha}{M^3 (\tilde{r}-2M)^2}\left[-2 \tilde{r}   \ln \left(1-\frac{2 M}{\tilde{r}}\right) +\frac{64 M^5}{\tilde{r}^4}-\frac{392 M^4}{\tilde{r}^3} +\frac{2512 M^3}{9 \tilde{r}^2}-\frac{4 M^2}{\tilde{r}}-4 M -\frac{4096 \tilde{r}}{945} \right. \nonumber \\
    &+\sqrt{\frac{\tilde{r}}{\tilde{r}-2 M}} \left(-\frac{8704 M^5}{27 \tilde{r}^4}+\frac{125824 M^4}{189 \tilde{r}^3} \left. -\frac{270848 M^3}{945 \tilde{r}^2}-\frac{2048 M^2}{945 \tilde{r}}-\frac{4096 M}{945}+\frac{4096 \tilde{r}}{945}\right) \right]. 
\end{align}
In this case, the weak-field limit yields
\begin{align}
    A(\tilde{r}) &= 1-\frac{2 M}{\tilde{r}}+\frac{16 \alpha  M^5 \epsilon }{9 \tilde{r}^9}+ \mathcal{O}\left(\frac{M^{10}}{\tilde{r}^{10}}\right), \label{eq:fT-weakSch-A-p3} \\
    B(\tilde{r}) &= 1+\frac{2 M}{\tilde{r}} +\frac{4 M^2}{\tilde{r}^2} +\frac{8 M^3}{\tilde{r}^3}+\frac{16 M^4}{r^4}+\frac{32 M^5}{\tilde{r}^5}+\frac{64 M^6}{\tilde{r}^6}+\frac{128 M^7}{\tilde{r}^7}+\frac{256 M^8}{\tilde{r}^8} +\frac{512 M^9}{\tilde{r}^9}-\frac{16 \alpha  M^5 \epsilon }{\tilde{r}^9} + \mathcal{O}\left(\frac{M^{10}}{\tilde{r}^{10}}\right). \label{eq:fT-weakSch-B-p3}
\end{align}
\end{widetext}

\begin{acknowledgements}
The authors would like to acknowledge the Cosmology@MALTA University of Malta award. AF is a recipient of the Institute of Space Sciences and Astronomy PhD scholarship award.
\end{acknowledgements}

\end{document}